# FRAMEWORK TO SOLVE LOAD BALANCING PROBLEM IN HETEROGENEOUS WEB SERVERS

Deepti Sharma[1] and Archana B.Saxena[2]

[1]Department of Computer Science, Jagan Institute of Management Studies, Affiliated to GGSIPU, Delhi
`deepti.jims@gmail.com`

[2]Department of Computer Science, Jagan Institute of Management Studies, Affiliated to GGSIPU, Delhi
`archanabsaxena@gmail.com`

## ABSTRACT

*For popular websites most important concern is to handle incoming load dynamically among web servers, so that they can respond to their client without any wait or failure. Different websites use different strategies to distribute load among web servers but most of the schemes concentrate on only one factor that is number of requests, but none of the schemes consider the point that different type of requests will require different level of processing efforts to answer, status record of all the web servers that are associated with one domain name and mechanism to handle a situation when one of the servers is not working. Therefore, there is a fundamental need to develop strategy for dynamic load allocation on web side. In this paper, an effort has been made to introduce a cluster based frame work to solve load distribution problem. This framework aims to distribute load among clusters on the basis of their operational capabilities. Moreover, the experimental results are shown with the help of example, algorithm and analysis of the algorithm.*

## 1. INTRODUCTION

Internet is becoming immeasurably popular all across the globe. With the ever increasing dependence on the Internet, the traffic on the World Wide Web has increased at an explosive rate [12]. In such a situation, there is a heavy pressure on web server in terms of performance, scalability and availability of web services. When number of requests increase from a particular site the response time from that website also increases, it's really important to properly handle the slow response time problem otherwise clients will get frustrated and will either refuse connection or will never visit that web site again. A website can handle slow response problem in variety of ways: by increasing server bandwidth (which is not feasible every time), using web proxy caching or mirror websites, answering only text and making use of monolithic or cluster web server. Among various solutions available to solve this slow response problem, cluster web server (when single domain name is served by more than one computer) is the best option.The most important concern of this strategy is to handle incoming load dynamically between all working

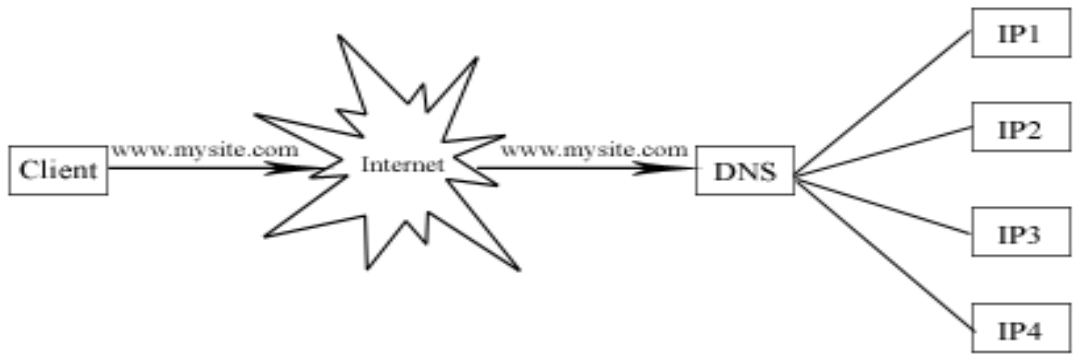

units.

Figure 1. End user or web client request (Domain Name) is directed to host sever through DNS. DNS stores IP address corresponding to Domain Names.

Handling load dynamically is a very big challenge but degree of challenge magnifies when we are working with cluster technology.

In this study, we propose load balancing technique framework which consists of a single control server and multiple heterogeneous clusters. Related work is reviewed in **section-2**. **Section-3**, formulates a problem in currently running strategies. **Section-4**, defines framework for new load balancing strategy for web server load balancing. **Section-5** illustrates the example. **Section-6** explains algorithm and **section-7** presents analysis of the proposed framework.

## 2. RELATED WORK

**Sandeep Sharma, Sarabjit Singh, and Meenakshi Sharma [2008]** have studied various Static(Round Robin and Randomized Algorithms, Central Manager Algorithm, Threshold Algorithm*)* and Dynamic (Central Queue Algorithm, Local Queue Algorithm) Load Balancing Algorithms in distributed system. The performance of these algorithms are measured by following parameters: Overload Rejection, Fault Tolerant, Forecasting Accuracy, Stability, Centralized or Decentralized, Nature of Load Balancing Algorithms, Cooperative, Process Migration,, Resource Utilization. Their result shows that static Load balancing algorithms are more stable with such parameters.

**Neeraj Nehra, R.B. Patel, V.K. Bhat [2007]** have proposed a DDLB(Dynamic Distributed Load Balancing) scheme for minimizing the average completion time of application running in parallel and improve the utilization of nodes. In proposed scheme instead of migrating the process for load balancing between clusters, they split the entire process into job and then balance the load. In order to achieve their target they will make use of MA (Mobile agent) to distribute load among nodes in a cluster.

**Helen D. Karatza and Ralph C. Hilzer [2002]** have worked together to find a load sharing policy in heterogeneous distributed environment where half of the total processor have double the speed of others except only two types of jobs: first class and generic. They have studied only non-preemptive job scheduling policies where scheduler have exact information about queue length of all processor and queuing time of dedicated jobs in fast processor.

**Andrew J. Page and Thomas J. Naughton** has proposed a genetic algorithm (GA) to dynamically schedule heterogeneous task on heterogeneous system in a distributed environment to minimize total execution time. GA uses historical information to exploit the best solution and completes it process in three steps: Selection, Crossover and Random mutations. The GA algorithm is only performed if there are more unscheduled tasks than processors;

**Kun-Ming V. Yu\*, Chih-Hsun Chou\* and Yao-Tien** Wang have designed and implemented a load balancing system based on fuzzy logic and proved that, this algorithm not only effectively reduces the amount of communication messages but also provides considerable improvement in overall performance such as short response times, high throughputs, and short turnaround times.

**Zhiling Lan, Valerie E. Taylor and Greg Bryan** have proposed a Load Balancing scheme for (SAMR) Structured Adaptive Mesh Refinement Application on distributed systems. In proposed scheme they consider the heterogeneity of processor and dynamic load on system and divide the complete Load balancing process into two phases Global Load Balancing and Local Load Balancing.

## 3. PROBLEM FORMULATION

The web servers of popular websites often need to be based on distributed or parallel architecture while preserving a virtual single interface. This will result into small latency time and less burden on each server. Different websites use different strategies to distribute load among web servers but most of the schemes concentrate on only one factor that is number of requests, but none of the schemes consider the point that:

- Different type of requests will require different level of processing efforts to answer.

- Status record of all the web servers that are associated with one domain name must be considered.

- Mechanism to handle a situation when one of the servers is not working.

## 4. SYSTEM MODEL / FRAMEWORK OF PROPOSED SCHEME

Proposed system [**Mi = {N, C, S}** where, **N** – Number of Heterogeneous Web Servers, **C**- Number of Clusters and **S** – Server Controller] can do following things:

- Server controller will build **'C'** clusters on the basis of Memory and CPU requirements.
- It will keep a record of status of all the machines (web server and clusters)
- When one of the web servers is not working, controller can monitor the faulty system.
- Uneven load distribution can be balanced within the cluster or between the clusters.
- It will work only for non executable jobs where none of the job can transfer between clusters during execution.

In this research, we have considered a model **'X'** where **'N'** (Heterogeneous) web servers form **'M'** clusters and a Server Controller **'S'** manages and regulates these **'M'** clusters. Every cluster is internally maintained and regulated by a controller **'C'.** These Cluster server **'C'** and Controller server **'S'** are responsible to distribute load and regulate centralized load balancing scheme within cluster and among the clusters respectively. To perform such activities they have to keep a record of assigned load, capabilities, processing speed and many more parameters with the help of: Review matrix, Ability Matrix and Load Matrix.

**Review Matrix**: This matrix is maintained within the Central Server Controller 'S' and within all Clusters 'C' which maintains the load conditions of every cluster and of every web server respectively at a particular point of time. It is the responsibility of every node (either cluster or web server) to update their Load status after every t! (t! is the fraction of seconds decided by 'S'). With this Load Matrix Server 'S' can make out whether the system is in load balanced state or it needs some load sharing between clusters. Similarly, Cluster 'C' can also decide about balanced or unbalanced state of web servers and can take actions accordingly.

**Ability Matrix:** This matrix is maintained by Server Controller **'S'** and Cluster Server **'C'**. Cluster Ability Matrix consists of information (Mac_Name, CPU speed, and Primary and Secondary memory details) about web servers that belong to that particular cluster. For Server controller Server Ability Matrix records information about cluster with a difference that server controller doesn't write exact value but ranges like: cluster1 consist of machines where CPU range can vary from 5000 to 10000 GHZ and so on for all other parameters. 'Server Ability Matrix' and 'Cluster Ability Matrix' are updated by Server controller and Cluster Server respectively and use at the time of load allocation.

**Load Matrix:** This matrix is maintained by Server Controller **'S'** and Cluster Server **'C'. 'S'** keeps a track of load status with each cluster and **'C'** is responsible to keep a track of load position with each web server within the cluster. It is responsibility of every web server to update load matrix of cluster server **'C'** after every **'T'** (fraction of time decided for updating load matrix) seconds and in case of controller servers **'S'** updating load matrix after decided time frame is the responsibility of cluster server **'C'**. Load Matrix helps Server Controller **'S'** and Cluster Server **'C '** in regulating load balanced condition with the cluster and among clusters.

This matrix also helps in identifying dead machines, after every **2*T!** (**T!** is the fraction of seconds decided by **'S'**) Load matrix is compared with Ability Matrix of Cluster 'Cluster Ability Matrix', any web server entry that is available in 'Cluster Ability Matrix' and no updates is received in last two update cycle then that machine is considered as dead machine or is assumed that some technical fault is there in that web server. Then Cluster Server **'C'** check Load matrix to find status of job (if any), that is assigned to dead machine, so that same job can be transferred to any other machine that is capable enough to process it.

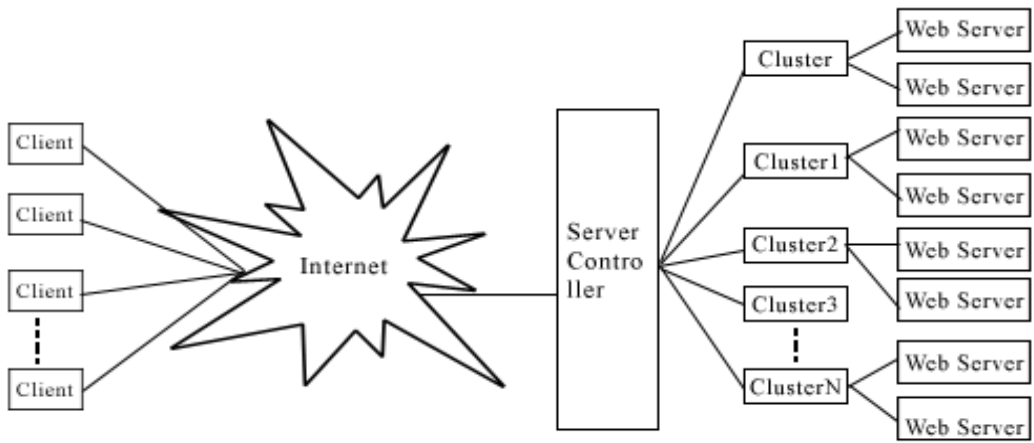

Figure 2. System architecture where all the web servers are divided into cluster on the basis of their operational capabilities and Server Controller is responsible to manage Load balancing and Load distribution among these clusters.

## 4.1 Working of Proposed System

With a view to implement above model, two strategies are defined:
- Job Assignment Strategy
- Load Balancing Strategy

**1) Job Assignment Strategy**

When a new http request *jig* is allocated to the system, it will be acknowledged by server controller **'S'**. Server controller will be responsible for distributing the load by effectively routing the user-requests over a clustered system. Server controller **'S'** does the following tasks:

- Searches Ability Matrix to find the cluster that can process the current job with best use of available resources.
- **jig** is transferred to the selected cluster.
- Then Cluster Servers **'C'** acknowledges the job receipt and searches (Ability Matrix) to find suitable web Server for executing in hand job.
- IP address of the web server is communicated to the client through **'C'** (cluster server) and **'S'** Server Controller for further communication between client and web server.

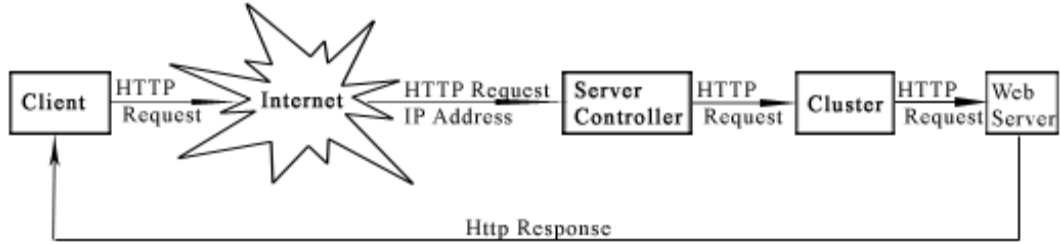

Figure 3. End user request is routed to Web Server through Server Controller & Cluster Server and further communication is done between web server and end user Client.

When a new job is presented to the system the same process is repeated.

**2) Load Balancing Strategy**

Apart from being an allocator, the next important role of 'S' and 'C' is to act as load balancer in the system. As a load balancer, they have to ensure a load balanced condition with in the cluster by two ways:

- Load balancing with-in cluster
- Load balancing between clusters

**4.1.1 Load balancing with-in cluster**

In this scheme, load distribution activity is initiated by the cluster server and proceeds with following actions:

➢ After every **T!** 'C' checks the Load matrix to find any uneven load condition then try balancing work among all available web servers.
➢ 'C' also checks for dead machine if any one and in Case one such machine is found then dead machine jobs are distributed among working machines.
➢ If none of the present cluster machines are able to share load, then the situation is updated to 'S' Server Controller.

**4.1.2 Load Balancing between clusters**

In this scheme, load distribution activity is initiated by the Server Controller **'S'** and proceeds with the following actions:

- After every **T! 'S'** checks Load matrix to identify any uneven work situation, ensures that load receiving cluster is able to process to the job the respective jobs and then balances the job allocations between clusters.

In both the Cases, information is updated in the Load Matrix of Cluster and Server Controller.

## 5. MATRIX FOR ALGORITHM

This section provides the example of the above proposed algorithm. Various matrixes are taken into consideration for evaluating the performance of proposed approach. As an example, seven jobs are being taken which will be distributed among two clusters with three web servers each. These are as follows:

### 5.1 Job Requirements Matrix

Table 1 illustrates the memory (in GB) and processing_speed (in MHz) requirements of seven jobs for execution.

Table 1. Job Requirements Matrix

| JOB_ID | Memory Requirement | Processing_Speed Requirement |
|---|---|---|
| 1 | 1500 | 80 |
| 2 | 500 | 50 |
| 3 | 1000 | 90 |
| 4 | 1200 | 40 |
| 5 | 600 | 50 |
| 6 | 700 | 60 |
| 7 | 500 | 60 |

### 5.2 Ability Matrix of Clusters

Cluster Ability Matrix consists of information (memory and Processing_speed) about web servers that belong to that particular cluster. Table 2 shows the ability matrix for Cluster 1 and Cluster 2. Thus,

WS1.1 is able to execute a job with memory requirement of 500 GB and processing speed of 60 MHz. Similarly, WS2.1 can execute a job which requires 1200 memory and 50 processing_speed and so on.

Table 2: Cluster Ability Matrix for Cluster 1 and Cluster 2

| WS_ID | MEMORY | PROCESSING_SPEED |
|---|---|---|
| WS1.1 | 500 | 60 |
| WS1.2 | 700 | 70 |
| WS1.3 | 1000 | 100 |

Cluster 1: (Range is between 0 and 1000)

| WS_ID | MEMORY | PROCESSING_SPEED |
|---|---|---|
| WS2.1 | 1200 | 50 |
| WS2.2 | 1500 | 70 |
| WS2.3 | 1800 | 80 |

Cluster 2: (Range is between 1000 and 2000)

### 5.3 Review Matrix of Servers

Table 3 shows review matrix for Cluster 1 and Cluster 2 before jobs assignment. Since, no jobs are assigned to any web server; memory_left is same as their initial memory.

Table 3: Review Matrix for Cluster 1 and Cluster 2 before Jobs assignment

| WS_ID | MEMORY | PROCESSING_SPEED | MEMORY_LEFT | JOBS_ASSIGNED | STATUS |
|---|---|---|---|---|---|
| WS1.1 | 500 | 60 | 500 | | EVEN |
| WS1.2 | 700 | 70 | 700 | | EVEN |
| WS1.3 | 1000 | 100 | 1000 | | EVEN |

Cluster 1

| WS_ID | MEMORY | PROCESSING_SPEED | MEMORY_LEFT | JOBS_ASSIGNED | STATUS |
|---|---|---|---|---|---|
| WS2.1 | 1200 | 50 | 1200 | | EVEN |
| WS2.2 | 1500 | 70 | 1500 | | EVEN |
| WS2.3 | 1800 | 80 | 1800 | | EVEN |

Cluster 2

Now, the seven jobs (shown in Table 1) have to be assigned among six web servers of Cluster 1 and 2. Table 4 shows the job assigned to all the web servers.

As shown in table 1, J1 requires 1500 memory and 80 processing speed. Firstly, the ability matrix of Cluster 1 is checked, but Cluster 1 memory range is in between 1 and 1000. So, the request will come to Cluster 2 and it will be fulfilled by WS2.3 (which is satisfying both memory and processing requirements). WS2.3 still has 300 memory left. Also the status is updated every time. Similarly, J2, J3, J4 and J5 jobs will also be assigned to WS1.1, WS1.3, WS2.1, and WS1.2 respectively. The next job i.e. J6 (with 700 of memory and 60 of speed is required), can be fulfilled by WS1.2 but this web server has only 100 GB of memory left. Thus the job will be left with memory of -600 and its load will be uneven. Similarly, when J7 will be assigned to WS1.1, its status will become uneven.

Table 4. Review Matrix for Cluster 1 and Cluster 2 after Jobs assignment

| WS_ID | MEMORY | PROCESSING_SPEED | MEMORY_LEFT | JOBS_ASSIGNED | STATUS |
|---|---|---|---|---|---|
| WS1.1 | 500 | 60 | -500 | J2+J7 | UNEVEN |
| WS1.2 | 700 | 70 | -600 | J5+J6 | UNEVEN |
| WS1.3 | 1000 | 100 | 0 | J3 | EVEN |

Cluster 1

| WS_ID | MEMORY | PROCESSING_SPEED | MEMORY_LEFT | JOBS_ASSIGNED | STATUS |
|---|---|---|---|---|---|
| WS2.1 | 1200 | 50 | 0 | J4 | EVEN |
| WS2.2 | 1500 | 70 | 1500 |  | EVEN |
| WS2.3 | 1800 | 80 | 300 | J1 | EVEN |

Cluster 2

### 5.4 Review Matrix after Load Balancing

As illustrated in our proposed approach, the algorithm will not only work as a job allocator but also as a load balancer. After every t! Second, the matrix will be checked and if there is any uneven load, it will be balanced among the other web servers. Balancing can be done within the cluster as well as between the clusters.

As seen in Table 4, the load status of WS1.1 and WS1.2 become uneven. In Cluster 1, only WS1.3 has even load status but there is no memory left for execution. If J6 or J7 will be assigned to it, this web server will become uneven. Thus load balancer will search for web server in Cluster 2. There is WS2.2 which is free and no job is being assigned to it. It can fulfill the requirements of both J6 and J7. Both the jobs will be assigned to it and thus load status of each web server will become EVEN. The resultant matrix is shown in Table 5.

Table 5: Review Matrix showing Load Status after Load Balancing

| WS_ID | MEMORY | PROCESSING_SPEED | MEMORY_LEFT | JOBS_ASSIGNED | STATUS |
|---|---|---|---|---|---|
| WS1.1 | 500 | 60 | 0 | J2 | EVEN |
| WS1.2 | 700 | 70 | 100 | J5 | EVEN |
| WS1.3 | 1000 | 100 | 0 | J3 | EVEN |

Cluster 1

| WS_ID | MEMORY | PROCESSING_SPEED | MEMORY_LEFT | JOBS_ASSIGNED | STATUS |
|---|---|---|---|---|---|
| WS2.1 | 1200 | 50 | 0 | J4 | EVEN |
| WS2.2 | 1200 | 70 | 300 | J6+J7 | EVEN |
| WS2.3 | 1800 | 80 | 300 | J1 | EVEN |

Cluster 2

## 6. ALGORITHMS FOR PROPOSED APPROACH

In this section, we define an algorithm for assignment and load balancing for above proposed approach.

### 6.1 Assignment Algorithm

This algorithm shows how a job is assigned within a cluster. There are three matrixes shown; **Job matrix** shows the list of jobs with their memory and processing speed requirement, **ability matrix** lists the server's memory and processing speed capabilities. Let there be n web servers WS in this cluster each with different memory and processing speed. This is maintained in the **review matrix** to see how many jobs have been assigned.

Now let there be a pool of m jobs for this particular cluster. The following algorithm also defines how a job is assigned to various cluster or web servers.

**Job Matrix: J** ( Job_ID, Memory and Processing Speed)
**Ability Matrix: A** (Server_ID, Memory and Processing Speed)
**Review Matrix: R** ( WS_ID, Processing_Speed, Memory_left, Jobs_Assigned and Status)
In this algorithm, there are two functions defined namely sort_servers ( ) and assignment ( ). Sort_servers ( ) sorts all the web servers within a cluster on the basis of its memory and processing speed. Assignment (A, J) assigns the job to the web servers. It first checks the cluster range and with in cluster it will search for the web server to which job can be assigned.

SORT_SERVERS ( )
{
  for(i=0 ; i<cluster[k].no._servers ; i++)
{
  for(j=i+1 ; j<cluster[k].no._servers ; j++)
{
  if(R[i][1] > R[j][1])
       Swap;
  else if(R[i][1] = = R[j][1])
   {    if(R[i][2]>R[j][2])
   {       Swap;
  } } } } }

ASSIGNMENT (A, J)
{
for (k=0 ; k<no_jobs ; k++)
{
for (i=0 ; i<no_cluster ; i++)
{ if (cluster[i].lower_memory < J[k].memory < cluster[i].upper_memory)
{  for (j=0 ; j<cluster[i].no_servers ; j++)
{ if (J[k].memory <= cluster[i].WS[j].memory&&J[k].processing <= cluster[i].WS[j].processing)
              ASSIGN (J[k])
     } } } } }

### 6.2 Load Balancing Algorithm

The following algorithm will show the load balancing by making use of review matrix and load matrix. This algorithm will consist of three functions namely sort ( ), load_balance ( ) and balance ( )

**REVIEW MATRIX: R (Server_ID, Present_Memory, Processing Speed and Load Status)**
**LOAD MATRIX: L (Cluster_ID, Memory Limit, Processing Limit and Load Stauts)**

**Sort** ( ) will sort the web servers in the Review matrix.

SORT (R)
{
  for(i=0;i<n;i++)
  {
   for(j=i+1;j<n;j++)
   {
       if(R[i][1] > R[j][1])
         { Swap; }
           else if(R[i][1] == R[j][1])
   {    if(R[i][2]>R[j][2])
           { Swap;}
  } } } }

**Load_Balance ( )** will check for load status of all web servers. If load status is uneven, it will call balance (cluster[i], job, j) to balance load within the cluster.

LOAD_BALANCE(R[n])

{

SORT(R);

for ( i=1 ; i <= n ; i++)

{

   for ( j=0; j < cluster[i].no_servers; j++)

  {

    if (load status = = uneven)

  {

  Flag=Call BALANCE (cluster[i] , job , j);//BALANCE WITHIN

      } } } }

**Balance ( )** will be executed for all web servers within the clusters as well as among the clusters. It will re-assign the uneven load status job to any of the web server where the job.memory and job.processing <= cluster.job and cluster.memory respectively.

If none of the web server load is even, balance ( ) will check it for rest of the clusters.

BALANCE (cluster[i] , job , j)

{

   for ( k = j ; k < cluster[i].no_servers ; k++)

  {

     if ( job.memory <= cluster[i].WS[k].memory &&

     job.processing <= cluster[i].WS[k].processing)

     {

        Assign job to this server and update R;

        Return(0);

     }

     else

      BALANCE(cluster[i++],job,0);  // BALANCE AMONG CLUSTERS

  } }

# 7. ANALYSIS OF THE PROPOSED ALGORITHM

This section illustrates the analysis of the above proposed algorithms.

To analyze the algorithms, let there be 'k' jobs, 'n' number of clusters and 'm' number of web servers. Thus, T(n) = Sorting + Assignment

**Sorting Algorithm:** Here for every i = 0 to m-1 there are m-1 to 1 comparisons i.e. (m-1) + (m-2) +---------------------2+1 = m (m-1)/2=$m^2/2$ comparisons. These comparisons are for a single cluster. So, **time complexity = T ($m^2/2$).**
For **'n' clusters**, the **time complexity = T ($nm^2/2$).**

## 7.1 Analysis for Assignment Algorithm

In this algorithm, the job is to be assigned to the web server which will match the memory and processing speed requirements of the job.

For every job we first match the memory requirements of the job with the cluster. If the cluster can serve the job then the comparison is done with the web servers WS within that cluster. This will avoid the unnecessary comparison of each job with every web server belonging to any cluster.
**Best Case:** When there is a single cluster with m web servers, there can be maximum m comparisons.
**Worst Case:** When there is n cluster with m web servers, there can be maximum $1.m_1+1.m_2$ -----------------+$1.m_n$ . If every cluster contains the same number of web servers say 'm' then the number of comparisons will be **'nm'**.
Following are given, some cases with different values of k (number of jobs), n (number of clusters) and m (number of web servers).

Let T (n) = sorting + assignment

**Case 1:** k=1, n=1, m=1
    Therefore, T(n) = T(0)+T(0);
    T(n) = 2T(0)
    T(n)=1

**Case 2:** k=1, n=1 and m>1

Therefore T(n)= T($m^2/2$)+T(m)
T(n)= O($m^2/2$)

**Case 3:** k=1, n>1 and m=1
Therefore T(n)= T(n) + T(n)
T(n) = O(n)

**Case 4:** k=1, n>1 and m>1

Therefore $T(n) = T(nm^2/2) + T(mn)$

$T(n) = O(nm^2/2)$

**Case 5:** k>1, n=1 and m=1

Therefore $T(n) = T(1) + T(k)$

$T(n) = O(k)$

**Case 6:** k>1, n=1 and m>1

Therefore $T(n) = T(m^2/2) + T(km)$

$T(n) = O(km^2/2)$

**Case 7:** k>1, n>1 and m=1

Therefore $T(n) = T(n) + T(kn)$

$T(n) = O(kn)$

**Case 8:** k>1, n>1 and m>1

Therefore $T(n) = T(nm^2/2) + T(knm)$

$T(n) = O(knm^2/2)$

## 7.2 Analysis for Load Balancing Algorithm

For Load balancing, first we sort all the web servers in each cluster on the basis of their memory left. Then, for balancing we check the status of the web server. If it is uneven, the load can be assigned to some other web server within the same cluster. If the web servers in the particular cluster are not able to take that load, it's passed on to the next cluster where it can be handled.

### 7.2.1 Load Balancing

For every cluster there will be 'm' steps to check the status of each web server and there are 'n' such clusters. If the web servers are balanced then there is no need of further processing and the **complexity remains T(nm).**

### 7.2.2 Load Balancing Within a Cluster

But if there are unbalanced nodes in the cluster then for every unbalanced nodes 'j' there will be (m-j) comparisons within the same cluster.

Let the number of unbalanced node = j. Following are the cases given with different values of n (number of clusters), m (number of web servers) and j (number of unbalanced jobs).

**Case 1:** When n>1, m>1 and j=0

$T(n) = T(nm^2/2)+T(nm)$

$T(n) = 2T(nm^2/2)$

$T(n) = O(nm^2/2)$, c=2

**Case 2:** When n>1, m>1 and j=1

T(n) = T(nm$^2$/2)+T(nm)+T(m-p), where p is the position of unbalanced node in the cluster

T(n) = 3 T(nm$^2$/2)

T(n)= O(nm$^2$/2) , c=3

**Case 3:** When n>1, m>1 and j>1

T(n)= T(nm$^2$/2)+T(nm)+T(j(m-p)) ,j<=m

T(n) = 3T(nm$^2$)

T(n) = O(nm$^2$) , c=3

### 7.2.3 Load Balancing among Clusters

If the load cannot be balanced within the same cluster then it is carried to another cluster. We make the comparisons there to get the best suitable web server that can handle the request.

So there are (n-i) clusters that can process a particular job as clusters are also arranged in ascending order of their memory limits. So, there will be at most (n-i)m comparisons in the worst case scenario.

**Case 1:** When n>1, m>1 and j=1

T(n) = T(nm$^2$/2)+T(nm)+T((m-p)+(n-i)m)

T(n) = T(nm$^2$/2)+2T(nm)

T(n) = 3T((nm$^2$/2)

T(n)= O(nm$^2$) , c=3

**Case 2:** When n>1, m>1 and j>1

T(n) = T(nm$^2$/2)+T(nm)+T(j((m-p)+(n-i)m))

T(n)= T(nm$^2$/2)+T(nm)+T(m$^2$+mn)

T(n)= T(nm$^2$/2)+T(nm)+T(m(m+n))

T(n)= T(nm$^2$/2)+2T(m(m+n))

T(n)=3T(m(m+n))

T(n)= O(m(m+n)) , c=3

## 8. CONCLUSIONS & FUTURE WORK

A fundamental merit of the proposed algorithm is its ability to trace dead machines. Further it has capability to divide and distribute web request on the basis of processing power involved. Our future work will focus on java implementation of this proposed algorithm and prove through simulation that this framework works well with heterogeneous web servers where incoming load is high and response is given within few seconds without any bottleneck.

One limitation left behind is suppose the main server controller fails then whole system will halt down. Because all requests are going through main server controller and if this server is unable to pass on the http request, framework will be shut down.

## REFERENCES

[1] Ezumalai R., Aghila G. and Rajalakshmi R. (2010), Design and Architecture for Efficient Load Balancing with Security Using Mobile Agents, IACSIT International Journal of Engineering and Technology Vol. 2, No.1, February, 2010 ISSN: 1793-8236.

[2] Sharma Sandeep, Singh Sarabjit and Sharma Meenakshi (2008), Performance Analysis of Load Balancing Algorithms, World Academy of Science, Engineering and Technology.

[3] Penmatsa, S., Chronopoulos A.T. (2007), "Dynamic Multi-User Load Balancing in Distributed Systems", IEEE International Symposium on Parallel and Distributed Processing, IPDPS'2007, ISBN: 1-4244-0910-1.

[4] Othman Ossama, and Schmidt Douglas C. (2007), "Optimizing Distributed System Performance via Adaptive Middleware Load Balancing", http://www.cs.wustl.edu/~schmidt/PDF/load_balancing_om_01.pdf .

[5] Bhat V.K., Nehra Neeraj, Patel R.B. (2007), "A Framework for Distributed Dynamic Load Balancing in Heterogeneous Cluster", Journal of Computer Science 3 (1): 14-24, 2007 ISSN 1549-3636.

[6] Hitz Markus A. and Mashraqi Farhan (2005), Cost-Effective Heterogeneous Web Clusters, 43rd ACM Southeast Conference, March 18-20, 2005, Kennesaw, GA, USA. Copyright 2005 ACM 1-59593-059-0/05/0003

[7] Krawczyk Henryk, Urbaniak Arkadiusz (2002), "Allocation Strategies of User Requests in Web Server Clusters", parelec, pp.217, International Conference on Parallel Computing in Electrical Engineering (PARELEC'02)

[8] Karatza Helen D. , Hilzer Ralph C. (2002), "Load Sharing In Heterogeneous Distributed Systems", Proceedings of 34th Winter Simulation Conference (WSC'02). ISBN:0-7803-7615-3.

[9] YM Teo and R Ayani (2001) , "Comparison of Load Balancing Strategies on Cluster-based Web Servers", Transactions of the Society for Modeling and Simulation.

[10] Aversa Luis, Bestavros Azer (2000),"Load Balancing a Cluster of Web Servers", Proceedings of IEEE International Performance, Computing, and Communications Conference (IPCCC'00), ISBN: 0-7803-5979-8.

[11] Malik S. (2000), "Dynamic Load Balancing in a Network of Workstation", 95.515 Research Report, 19 November, 2000.

[12] Rishi Khasgiwale, "Dispatcher-based Transparent-Dynamic Load Balancing algorithm for Web server clusters", Department of Electrical and Computer Engineering, University of Massachusetts-Amherst, MA – 01002, As a part of curriculum project for ECE-697A.
 www-nix.ecs.umass.edu/~rkhasgiw/.../rishi_report_ece697a.pdf.

**Authors**

Deepti Sharma is an Asst. Professor in Department of Computer Science at Jagan Institute of Management Studies (Affiliated to Guru Gobind Singh Indraprastha University), Rohini, Delhi. She is MPhil, MCA and pursuing her PhD (Computer Science). She has more than seven years of teaching experience in the areas of C, C++, Data Structures, Operating System and DBMS. Her research areas include "Load Balancing in Hetrogenous Web Servers" and "Mobile Banking" on which papers have been published in National and International conferences and journals. Various seminars, workshops and FDP (AICTE) have been attended.

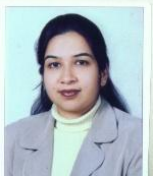

Archana B Saxena is working as Assistant Professor and Web Administrator in Jagan Institute of Management Studies (JIMS) since Oct 2004. Presently she is perusing PhD in Computer Science from GNDU on topic "Load Balancing in Heterogeneous Distributed Systems". She has attained MPhil Degree in CS from MKU, in 2006, MCA from IGNOU in 2003 and Bachelors in Commerce from Delhi University. She has served as guest Lecturer in ARSD, Delhi University and Faculty as NIIT. She has written various papers on Load Balancing, Mobile Computing and Data Mining published in various journals and conferences.

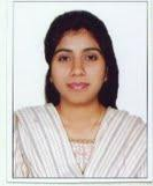